\documentclass[]{article}
\pdfoutput=1 
\usepackage{jheppub}
\usepackage[T1]{fontenc} 
\usepackage{color}
\usepackage{mathtools}
\usepackage[caption = false]{subfig}
\usepackage[normalem]{ulem}

\newcommand{\eq}[1]{\begin{equation}\begin{split} #1 \end{split}\end{equation}}
\newcommand{\eqs}[1]{\begin{align} #1 \end{align}}

\usepackage{graphicx}
\graphicspath{ {Plots/} }

\usepackage{circuitikz}
\usepackage{tikz}
\usepackage{rotating}
\usetikzlibrary{arrows}
\usetikzlibrary{calc}
\usetikzlibrary{decorations}

\preprint{UTTG--09--17}

\title{Holographic Complexity and Noncommutative Gauge Theory}

\author[b]{Josiah Couch,}
\author[b]{Stefan Eccles}
\author[b]{Willy Fischler}
\author[b]{and     Ming-Lei Xiao}

\affiliation[b]{Theory Group, Department of Physics and Texas Cosmology Center, The University of Texas at Austin, Austin, TX 78712, USA}

\emailAdd{josiah.couch@utexas.edu}
\emailAdd{stefan.eccles@utexas.edu }
\emailAdd{fischler@physics.utexas.edu}
\emailAdd{mingleix@utexas.edu}

\abstract{We study the holographic complexity of noncommutative field theories. 
The four-dimensional $\mathcal{N}=4$ noncommutative super Yang-Mills theory with Moyal algebra along two of the spatial directions has a well known holographic dual as a type IIB supergravity theory with a stack of D3 branes and non-trivial NS-NS B fields. We start from this example and find that the late time holographic complexity growth rate, based on the "complexity equals action" conjecture, experiences an enhancement when the non-commutativity is turned on. This enhancement saturates a new limit which is exactly 1/4 larger than the commutative value. 
We then attempt to give a quantum mechanics explanation of the enhancement. 
Finite time behavior of the complexity growth rate is also studied. 
Inspired by the non-trivial result, we move on to more general setup in string theory where we have a stack of D$p$ branes and also turn on the B field. Multiple noncommutative directions are considered in higher $p$ cases. }

\begin{document}
\maketitle

\section{Introduction}
\label{sec:intro}

In the past several years, there has been a growing interest in the topic of "holographic complexity." 
This interest was originally motivated by the late time growth of the wormhole volume in two sided black holes, which seems to have no correspondence in the boundary which is in thermal equilibrium. It was then conjectured that such a phenomenon should be related to the quantum complexity of the boundary state \cite{Susskind:2014rva}, and this conjecture was strengthened by study of quantum chaos, namely the "switchback effect" \cite{Stanford:2014jda, Susskind:2014jwa}. 
There have since been several conjectures as to the exact quantity dual to complexity on the boundary, all tied to the phenomenon of expanding wormholes in two-sided black holes. The first proposal was that complexity is dual to the volume of a maximal spatial slice with a given boundary \cite{Susskind:2014rva}, and the next \cite{Brown:2015bva, Brown:2015lvg} was the gravitational action evaluated on the Wheeler-DeWitt (WDW) patch. A third closely related conjecture was later proposed in \cite{Couch:2016exn}, namely that the complexity is dual to the space-time volume of a WDW patch. 

Unfortunately, there is little that we know about the concept of quantum complexity in the boundary field theory. 
The basic definition involves a reference state $|\psi_0\rangle$, a set of quantum gates $G = \{ g_i\}$, and a tolerance parameter $\epsilon$. The complexity of a quantum state $|\psi\rangle$ is the minimum number of gates one needs to make up a quantum circuit $Q = \prod_{i=1}^\mathcal{C} g_i$ so that $d_f(Q|\psi_0\rangle , |\psi\rangle ) < \epsilon$. One can also define the complexity of a unitary operator $U$ to be the minimum number of gates one needs to make up a quantum circuit $Q_U$ so that $|| Q_U - U || < \epsilon$. 
\footnote{$d_f(\ ,\ )$ is the Fubini-Study metric for quantum state $d_f(\alpha,\beta) = \arccos\sqrt{\dfrac{|\langle\alpha|\beta\rangle|^2}{\langle\alpha|\alpha\rangle\langle\beta|\beta\rangle}}$. The norm $||A||$ for operators can be defined as the square root of the spectral radius $\rho(A^\dagger A)$, which is the supremum of the eigenvalues of $A^\dagger A$.}
The holographic complexity is supposed to be the state complexity, while we also use the operator complexity to analyze the characteristic behavior in section \ref{sec:3}. 
Even with these definitions, the task of actually computing the relative complexity of two states is notoriously difficult. What is more, in the definition one has to make several choices, and where these choices appear in the holographic prescription is as of yet unclear. It is also a puzzle how one goes from the discretum of quantum circuits to a supposedly continuous quantum field theory. There has been considerable effort defining complexity in the quantum field theory \cite{Hashimoto:2017fga, Jefferson:2017sdb, Caputa:2017yrh, Czech:2017ryf, Chapman:2017rqy, Yang:2017nfn}, however they are weakly related to the holographic complexity at this point. 
%
%
Therefore, what we are interested in is to utilize our intuitions from quantum mechanics to conjecture some constraints on complexity in general. These constraints are to be tested for both the boundary theory and the holographic theory. 

Among the constraints which people have considered is the Lloyd bound \cite{Lloyd}. This bound was derived from the Margolus-Levitin theorem \cite{Margolus:1997ih} under the assumption that each gate will evolve a generic state into an orthogonal state. It states that the time rate change of complexity\footnote{We also refer to the time rate change of complexity as the ``complexification" rate, which should be considered synonymous as they appear in this paper.} is constrained by the energy:
\eq{\label{eq:bound}
\dot{\mathcal{C}} \leq \frac{2M}{\pi},
}
where $M$ is the energy of the system. 
In \cite{Brown:2015bva, Brown:2015lvg} it was conjectured that neutral black holes should saturate this bound, and this assumption was made in order to set the constant of proportionality between complexity and action. 
This conjecture originated from the fast scrambling nature of black holes and the related idea that black holes are the fastest possible quantum computers. However, one finds that for neutral black holes, the Lloyd bound is saturated from above \cite{Carmi:2017jqz}, which makes the conjecture somewhat suspicious. 
One can also argue that the Lloyd bound is not an exact bound because the assumption is based on is highly unrealistic. In fact, whether this assumption applies in the case of holographic complexity has recently been questioned in \cite{Cottrell:2017ayj}.

In light of these difficulties with the Lloyd bound, it is interesting to test the holographic complexity conjectures\footnote{In this paper we will consider only complexity = action, and discussion of the complexity = volume and complexity = spacetime volume conjectures are left for future work.}
against additional pieces of intuition in novel contexts. One context which might reasonably provide a testbed is the noncommutative field theories. 
The study of such theories has a long history and has produced many profound results, see for example \cite{Seiberg:1999vs,Maldacena:1999mh,Hashimoto:1999ut,Cai:1999aw,Alishahiha:1999ci,Berman:2000jw}. One feature of noncommutative field theory which is suggestive of interesting behavior is that it adds a degree of non-locality, which has been shown to lead to interesting effects, e.g. an increase relative to the commutative case in the dissipation rate of scalar modes \cite{Edalati:2012jj}. 
Indeed, the holographic entanglement entropy in this context has already been studied in, for example, \cite{Fischler:2013gsa, Karczmarek:2013xxa}, where non-trivial behavior was found in the limit where the Moyal scale is much larger than the thermal scale. 
The geometry was obtained in a string theory context by turning on the NS-NS B fields on Dp branes. The non-vanishing B field then induces Dirichlet boundary condition for open strings, and non-zero commutator of the end point coordinates \cite{Seiberg:1999vs}. After decoupling the closed strings, the Dp brane worldvolume becomes a noncommutative space. It was shown that in such setup, although space is coarse-grained by the Moyal scale, which might indicate a reduction in the number of degrees of freedom, it turns out that all thermodynamical quantities are unchanged \cite{Cai:1999aw, Maldacena:1999mh}. 
This can be understood by looking at the thermal boundary state in the large N limit, which consists of only planar diagrams without external legs. Such diagrams are insensitive to the non-commutativity of the spacetime \cite{Fischler:2000bp}. 
It thus provides a perfect arena for testing quantum complexity, whose main characteristic is that it is more than thermodynamics. If the holographic complexity can see the difference caused by non-commutativity, it is a sign that we are on the right track.

The remainder of this paper is organized as follows: 
In section \ref{sec:2} we construct the holographic dual of a noncommutative super Yang-Mills (NCSYM) theory and compute the holographic complexity of a state on the boundary using the CA proposal. The complexity growth rate is given as a function of the Moyal scale $a$, the horizon radius $r_H$ and time $t$, and at late times its monotonic enhancement with $a$ is shown. 
In section \ref{sec:3}, we attempt to give a quantum mechanical explanation of the enhancement of late time complexity growth rate. 
In section \ref{sec:4}, we discuss the finite time behavior of our result and compare to the recent independent studies \cite{Carmi:2017jqz}. 
To make our result more convincing, we explore more examples with non-commutativity in section \ref{sec:5}. We have a similar setup as in section \ref{sec:2} in various dimensions and we have various numbers of pairs of noncommutative directions.
In \ref{sec:Conclusion}, we conclude with a brief discussion of our results and make a few remarks of possible directions for future studies. 
In the appendix \ref{app:1}, we show the explicit calculation for the WDW patch action. 
Appendix \ref{app:3} talks about the thermodynamic property of the D$p$ brane solutions.



\section{Holographic Complexity of 4d $\mathcal{N}=4$ NCSYM}
\label{sec:2}

\subsection{The holographic dual to noncommutative SYM}
\label{subsec:dualsystem}

We consider the noncommutative field theory widely studied in the context of string theory. It was shown that the non-vanishing NS-NS B field will induce noncommutative space on the D brane that decouples from the closed string excitations \cite{Seiberg:1999vs}. The way to turn on the B field is to perform a T duality, in D3 brane for instance, along $x_3$ direction, assuming the $x_2, x_3$ are compatified on a torus. The torus becomes tilted after the T duality, which indicates a D2 brane smearing along $x_3$ direction. Then one performs another T duality along $x_3$, to get the following solution (\cite{Hashimoto:1999ut, Maldacena:1999mh}):

\eqs{
\begin{split}
& ds^2 = \alpha' \bigg[ \left(\frac{r}{R}\right)^2
\left(- f(r) dt^2 + dx_1^2  + h(r) (dx_2^2 + dx_3^2)\right)
+\left(\frac{R}{r}\right)^2 \left(
\frac{dr^2}{f(r)} + r^2 d\Omega_5^2\right)\bigg], \\
& f(r) = 1 - \left(\frac{r_H}{r}\right)^4 ,\quad h(r) = \frac{1}{1 + a^4 r^4}, \\
\end{split} \\
\begin{split}
\label{eq:field solutions}
& e^{2\Phi} = \hat{g}_s^2 h(r) ,\\
& B_{23} = B_{\infty}(1-h(r)) ,\quad B_{\infty} = -\frac{\alpha'}{a^2R^2} , \\
& C_{01} = -\frac{\alpha^{\prime} a^2 r^4}{\hat{g}_s R^2} ,\quad F_{0123r} = \frac{4\alpha^{\prime 2} r^3}{\hat{g}_s R^4} h(r).
\end{split}
}
The $\{ t, x_1, x_2, x_3\}$ are the D3 brane coordinates, while $\{ x_2, x_3\}$ are non-commuting with Moyal algebra
\eq{
[x_2, x_3] = ia^2.
}
The radius coordinate $r$ has units of inverse length\footnote{In the literature, the coordinate denoted here by 'r' is typically denoted 'u' in order to emphasize that it does not have dimensions of length. We have however chosen to denote it by 'r' to avoid confusion with the Eddington-Finkelstein like null coordinate}, and $a$ is the Moyal scale with units of length. $r_H$ denotes the location of the event horizon, and $\hat{g}_s$ denotes the closed string coupling, which is related to the $S^5$ radius as $R^4 = \hat{g}_sN$.

Note that the geometry becomes degenerate at $r\to\infty$; thus we have to put the boundary theory on some cutoff surface $r_b < \infty$. It was shown that this natural cutoff plays an important role in the divergent structure of entanglement entropy \cite{Fischler:2013gsa}. However, as will be explained later, our computation is cutoff independent; therefore we don't need to worry about it.

As explained in \cite{Maldacena:1999mh}, all the thermodynamic quantities of this solution are the same as in the commutative case. In particular, the temperature and entropy is independent of $a$, given by
\eq{\label{eq:p3thermo}
&E = \frac{3r_H^4\Omega_5 V_3}{(2\pi)^7\hat{g}_s^2 }\\
&T = \frac{r_H}{\pi R^2}, \\
&S = \frac{4\pi R^2 r_H^3\Omega_5 V_3}{(2\pi)^7\hat{g}_s^2}\\
}
It is then interesting to ask whether the complexity is affected by the non-commutativity because complexity is fine-grained information that knows more than thermodynamics. 

We adopt the Complexity equals Action (CA) approach to compute the holographic complexity of the boundary state. It involves evaluating the action in a bulk subregion, called the Wheeler-deWitt (WDW) patch. Recent work on evaluating gravitational action \cite{Lehner:2016vdi} provided a toolkit that deals with null boundary contributions in the context of Einstein gravity. Hence we are interested in the Einstein frame action of type IIB supergravity:
\eqs{
& ds_E^2 = \exp(-\Phi/2)ds^2, \label{metric_E}\\
& 2\kappa^2 S_E = \int d^{10} x \sqrt{-g_E} \left[ \mathcal{R} - \frac{1}{2} |d\Phi|^2 - \frac{1}{2} e^{-\Phi} |dB|^2 - \frac{1}{2} e^{\Phi} |F_3|^2 -\frac{1}{4}|\tilde{F}_5|^2 \right]
-\frac{1}{2}\int C_4 \wedge dB \wedge F_3 \label{eq:IIB-action},
} 
where the notation $|F_p|^2 = \frac{1}{p!}F_{\mu_1...\mu_p}F^{\mu_1...\mu_p}$ is understood. One should keep in mind that the 5-form $\tilde{F}_5$ is self dual while evaluating this action. This requirement actually always makes the term $|\tilde{F}_5|^2 = 0$.
\footnote{We point it out that due to the famous subtlety about type IIB action, that the self-duality condition should be imposed by hand, the treatment we use for the action is only plausible. There are other ways to impose self-duality, for example the PST formulation, but the action computation and the holography there will be subtle.}

%
%
%
%
%
%

\subsection{Wheeler-DeWitt Patch Action}

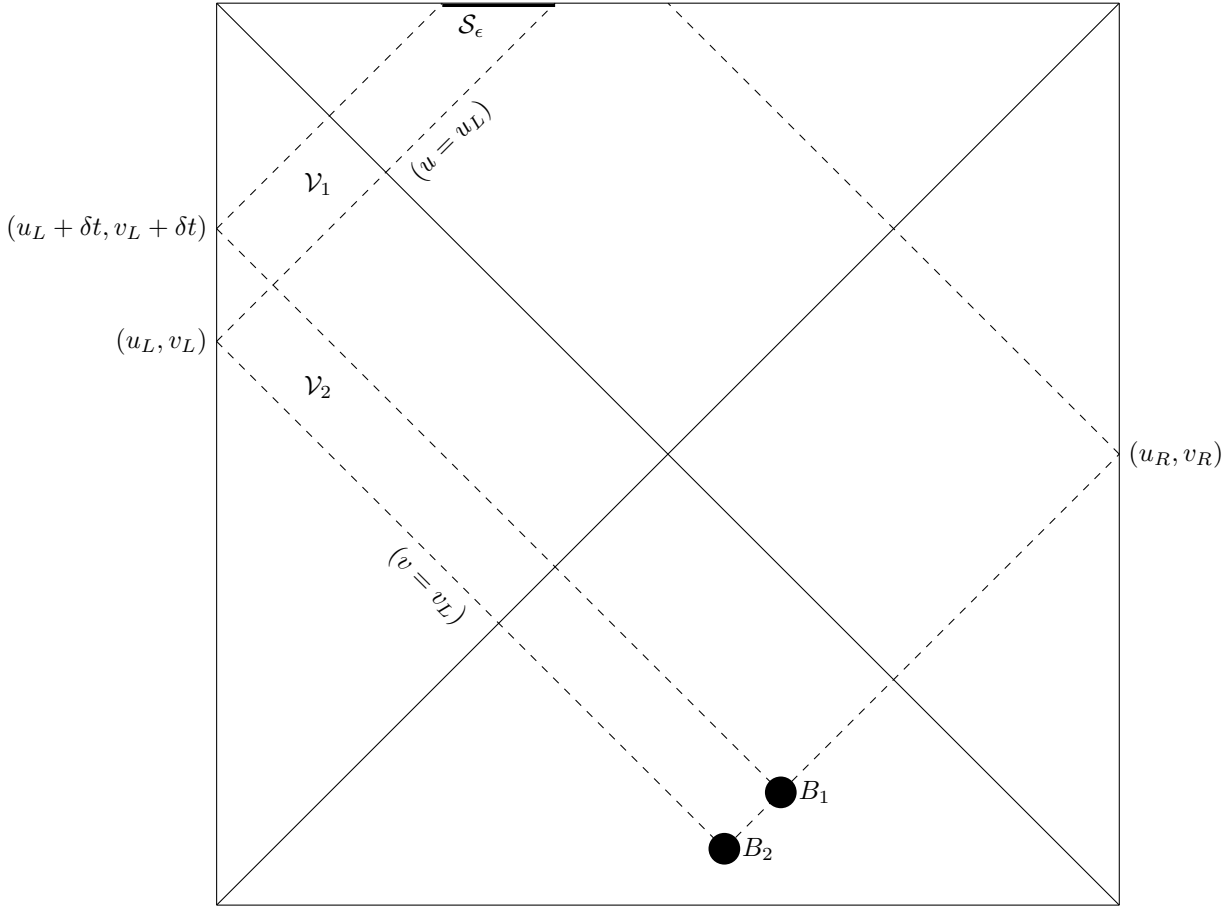
\begin{figure}
    \begin{center}
        \begin{tikzpicture}[scale=3]
        \draw (0,0)--(4,0);
        \draw (4,0) --(4,4);
        \draw (4,4)--(0,4);
        \draw (0,4)--(0,0);
        \draw (0,0) --(4,4);
        \draw (0,4)--(4,0);
        \draw[dashed] (0,3)--(1,4);
        \draw[very thick] (1,3.99)--(1.5,3.99);
        \draw[dashed] (0,2.5)--(1.5,4); 
        \draw[dashed] (0,3)--(2.5,.5); 
        \draw[dashed] (0,2.5)--(2.25,.25);
        \draw[dashed] (2.25,.25)--(4,2);
        \draw[dashed] (4,2)--(2,4);
        \fill(2.5,.5) circle[radius=2pt];
        \fill(2.25,.25) circle[radius=2pt];
        \draw (0.45,3.2) node{$\mathcal{V}_1$};
        \draw (0.45,2.3) node{$\mathcal{V}_2$};
        \draw (2.4,.25) node{$B_2$};
        \draw (2.65,.5) node{$B_1$};
        \draw (1.13,3.9) node{$\mathcal{S}_\epsilon$};
        \draw (0,2.5) node[left]{$(u_L,v_L)$};
        \draw (0,3) node[left]{$(u_L+\delta t,v_L + \delta t)$};
        \draw (4,2) node[right]{$(u_R,v_R)$};
        \draw (.75,1.6) node[right, rotate=-45]{$(v=v_L)$};
        \draw (.85,3.2) node[right, rotate=45]{$(u=u_L)$};
        \end{tikzpicture}
    \end{center}
    \caption{Two WDW patches separated by $\delta t$.  Although the boundary of each patch is really at some large but finite $r_b$, the choice of $r_b$ drops out in the differences we consider and we do not indicate it explicitly in this graphic.}
    \label{fig:WDW}
\end{figure}

The WDW patch is defined to be the union of all spatial slices anchored on a boundary time slice $\Sigma$. Regarding representing the boundary state, the WDW patch differs from the entanglement wedge at two points: first, it specifies a specific time slice on the boundary, instead of a covariant causal diamond; second, it probes behind the horizon, which is supposed to contain information beyond thermodynamics. It was conjectured in \cite{Brown:2015bva, Brown:2015lvg} that the action evaluated in the WDW patch is dual to the relative complexity of the quantum state living on $\Sigma$. This conjecture is referred to as `complexity = action' or CA duality. 
In our noncommutative geometry setup, we will be interested in the WDW patch for the two-sided black hole, which intersects the left boundary at time $t_L$, and the right boundary at time $t_R$. According to CA quality, the action evaluated on such a patch will compute the relative complexity of the quantum state of the boundary CFT living on the ($t_L$, $t_R$) slice as
\eq{\label{eq:prop_k}
\mathcal{C}(t_L,t_R) = k S_{WDW},
}
with the coefficient set to $k^{-1} = \pi\hbar$ by the assumption that AdS-Schwartzchild black hole saturates the Lloyd bound. The complexity computed this way is cutoff dependent, but its time derivative
\begin{equation}\label{eq:timederivative}
\dot{\mathcal{C}}(t_L,t_R) := \frac{d}{dt_L} \mathcal{C}(t_L,t_R),
\end{equation}
in which we are interested, is cutoff independent. Notice that our choice to differentiate with respect to the left time is arbitrary, as the geometry should be symmetric between left and right.
It will prove convenient to utilize radial advanced/retarded null coordinates:
\begin{equation}
dr^* =\frac{R^2dr}{r^2f(r)},\quad u=t+r^*,\quad v=t-r^* .
\end{equation}
Notice that unlike $r$, $r^*$ has units of length. 
Suppressing all but the bulk and timelike direction, the contributions to the time rate change of the WDW patch can be visualized in the conformal diagram represented in Figure \ref{fig:WDW}.

The calculation of the time rate change of the action is detailed in Appendix \ref{app:1}. It is convenient to express the result in terms of the radial coordinate $r_B$ of the pastmost joint of the WDW patch (joint $B_2$ in the diagram \ref{fig:WDW}, which coincides with joint $B_1$ as $\delta t \rightarrow 0$.)  Note that $r_B$ increases monotonically with $t_L$ from $r_B=0$ to $r_B=r_H$ as $t_L\rightarrow \infty$, and so we will use it to parameterize the time dependence of the complexification rate.
\footnote{We consider only $t_L>0$, and fix $t_R$ so that this corresponds to when the joint $B$ has left the past singularity.}
We find the following combined result:
\begin{equation}\label{eq:finiteTimeRate}\begin{split}
\dot{S}_{WDW}&=\frac{ \Omega_5 V_3}{(2\pi)^7\hat{g}_s^2}
\bigg(\frac{-2\log(1+a^4r_B^4)}{a^4}+4r_B^4+6r_H^4+3(r_H^4-r_B^4)\log\big|\frac{c \bar{c}\sqrt{\hat{g}_s}R^2 r_B^2}{\alpha (1+a^4 r_B^4)^{1/4}(r_H^4-r_B^4)}\big|\bigg)
\end{split}\end{equation} 
where $c$ and $\bar{c}$ are arbitrary constants associated  with the normalization of boundary null generators entering the computation of $\delta S_{\text{joint}}$.  See Appendix \ref{appsubsec:joints}, as well as \cite{Lehner:2016vdi}, \cite{Jefferson:2017sdb} for discussion.

Various aspects of the time dependence (or $r_B$ dependence) of equation \ref{eq:finiteTimeRate} are unusual in light of the conjectured CA duality.  Similar features have been seen in other systems \cite{Carmi:2017jqz}.  We discuss the finite time behavior in Section \ref{sec:4}.

The late time complexification rate is achieved by sending $r_B\rightarrow r_H$:
\begin{equation}
\dot{S}\big|_{t\rightarrow \infty}\approx \frac{\Omega_5 V_3 r_H^4}{(2\pi)^7\hat{g}_s^2}\left(10-2\frac{\log(1+a^4r_H^4)}{a^4r_H^4}\right)
\end{equation}

One can immediately see that if we assume the standard relationship, $\mathcal{C}=k S$ with $k=1/ \pi$, then the system violates the Lloyd bound (\ref{eq:bound}) at late times:  the ratio $\frac{\dot{S}}{2M}$ should be less than or equal to 1, but at late times it saturates values between 4/3 to 5/3 as we vary $a$.  The relevance of the bound to holographic complexity has been disputed \cite{Cottrell:2017ayj}, and violations have been found in many other systems. But for purposes of comparison we find it interesting that, even if we had not assumed the standard $k=1/\pi$, but instead used the logic that commutative black holes should saturate the Lloyd bound, we would set $k=3/(4\pi)$.  Clearly, the associated bound would fail immediately upon considering highly noncommutative black holes.  Rather than proposing some different $k$ in the relationship $\mathcal{C}=k S$, we find it plausible that such a choice does not generalize to all systems, at least under the current conventions for computing bulk action.

Overlooking the Lloyd bound for now, the dependence of the late time complexification on the noncommutativity parameter $a$ is rather striking.

\begin{figure}[htbp]
    \begin{center}
        \includegraphics[scale=0.7]{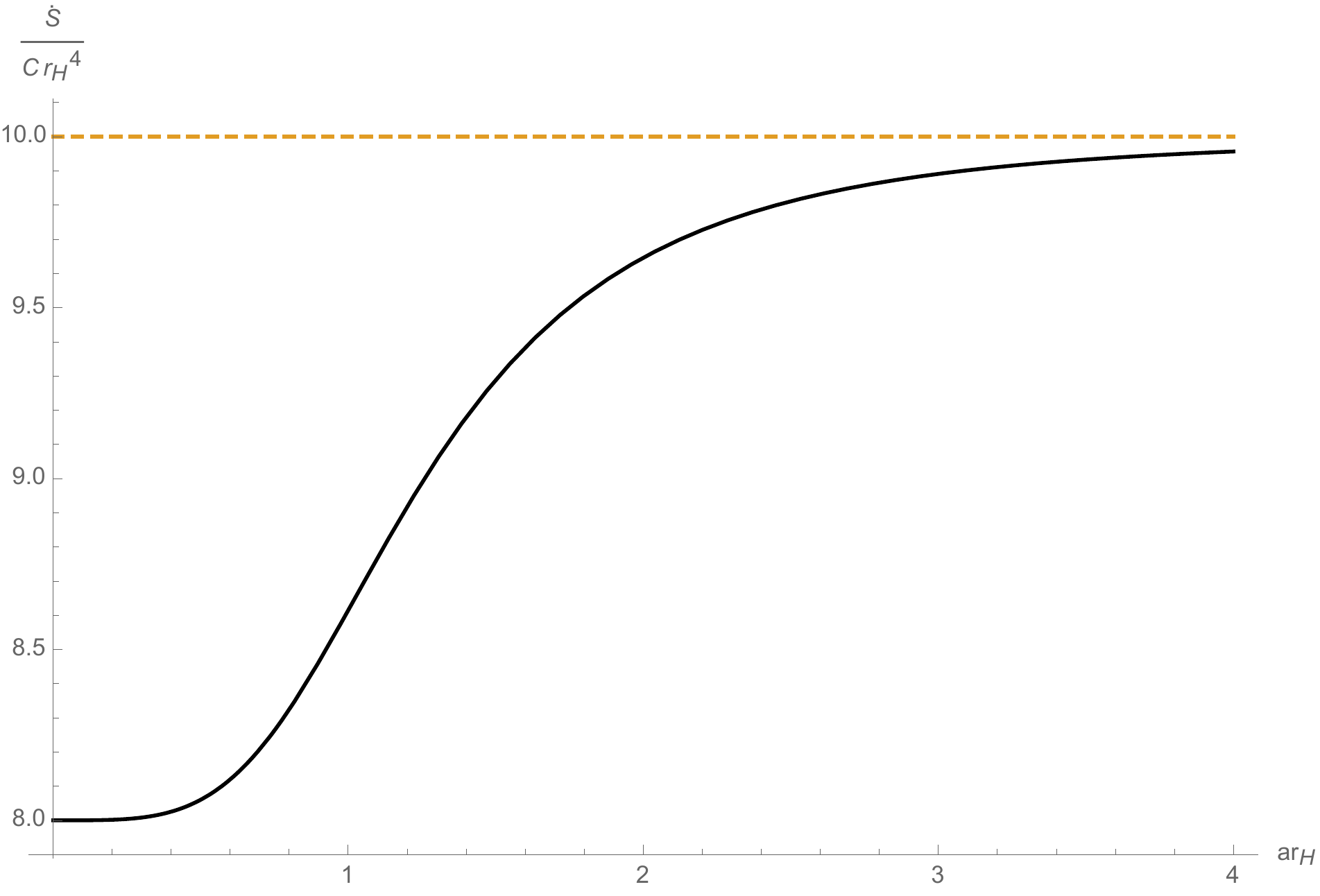}
    \end{center}
    \caption{Late time action growth rate normalized by $C=\frac{\alpha^4 \Omega_5 V_3}{\hat{g}_s^2}$ and extra $r_H$ dependence, versus $a r_H$, which is the Moyal scale measured in units of thermal length. It is observed that the complexification rate under the CA conjecture increases significantly when the Moyal scale is comparable to the thermal scale, and saturate a new bound which is 5/4 of the commutative value when the Moyal scale is much larger than the thermal scale.} 
    \label{fig:LateTime}
\end{figure}

As one can see from Figure \ref{fig:LateTime}, the complexification rate increases with the non-commutativity parameter $a$, or more specifically the Moyal scale. It's also intriguing that $a$ always appears in the combination $ar_H$, indicating that the only reference scale in the theory that the Moyal scale is sensitive to is the thermal scale $T^{-1} \sim r_H^{-1}$. When $a\ll T^{-1}$, the complexification rate does not change much. It noticeably changes when $a$ becomes comparable to $T^{-1}$. When $a \gg T^{-1}$, the complexification rate stops growing and saturates a new bound. It is inspiring to see that it does not grow indefinitely because that will violate the Lloyd bound in any possible sense. On the other hand, the ratio that it increases is an interesting rational number 5/4. It may imply that this enhancement could be understood as some counting problem. With all these interesting features in mind, we want to answer two questions: 
\begin{enumerate}
    \item How might we explain the enhancement from non-commutativity? 
    
    \item Are there other examples of noncommutative theories that corroborate these results?
\end{enumerate}
These will provide the content for the next few sections.

%
%



\section{Non-Commutativity Enhancement of Complexification Rate}
\label{sec:3}

Why the above enhancement should be exactly 25\% is as of yet unclear. We do, however, have a conceptual argument for why there should be a noncommutative enhancement at all. 

Consider the following problem: We have a unitary operator $U$, whose complexity is known to be $\mathcal{C}(U)$, and we want to know what can be said about the complexity of $\mathcal{C}(U^N)$ for some integer $N$. One thing that can be immediately said is that

\begin{equation}\label{Bound}
\mathcal{C}(U^N) \leq N \mathcal{C}(U)
\end{equation}\label{eq:Bound}

Because given an optimal circuit $Q$ implementing $U$, $U^N$ can be implemented by $N$ successive applications of $Q$, namely $Q^N$.
\footnote{There is a subtlety here in that $Q$ only need implement a unitary that is within some small number $\epsilon$ of $U$, but if this is the case, there is no guarantee that $Q^N$ will be within $\epsilon$ of $U^N$. It is also possible that for particular choices of gate set, some power of $Q$, say $Q^M$, may itself be a gate.  This would result in ``saw tooth" growth in complexity and periodically discontinuous time derivatives. It may be hoped that such issues are rendered obsolete in an appropriate continuum limit (as in the ``geometry of complexity" program \cite{Brown:2017jil, Brown:2016wib}), and we ignore these subtleties for the present discussion.}
The bound above need not be saturated, however, as there might be a few gates at the beginning of $Q$ which can cancel with some at the end of a successive copy of $Q$, resulting in a new circuit which (a unitary identical) to $Q^N$, but which is less complex. If we suppose that every time a new copy of $U$ is added (after the first one of course), we get a cancellation of $\chi$ gates, and we suppose that $\chi$ doesn't depend on $N$ (or at least asymptotes to a constant as $N$ becomes large), then we have
\begin{equation}
\mathcal{C}(U^N) \approx N \mathcal{C}(U) - (N-1) \chi
\end{equation} 
It's easy to show that this formula holds for any $U \to U^n$ with the same $\chi$.

If we are then interested in the (time evolution of the complexity of a family of operators) generated by some hamiltonian $H$
\begin{equation}
U(t) = e^{i H t},
\end{equation}
then we may use the above to write
\begin{equation}\label{eq:rateansatz}
\mathcal{C}(t) \equiv \mathcal{C}(U(t)) = \mathcal{C}[U(\delta t)^{t/ \delta t}] \approx \frac{t}{\delta t} \left[ \mathcal{C}(\delta t) - \chi \right] + \chi.
\end{equation}
This will be true for any $t$ and $\delta t$.
%
%
Therefore we can compute the complexification rate
\begin{equation}
\frac{d}{dt} \mathcal{C}(t) \approx \frac{1}{\delta t} \left[\mathcal{C}(\delta t) - \chi \right].
\end{equation}

\begin{figure}
    \begin{center}
        \includegraphics[scale=1.2]{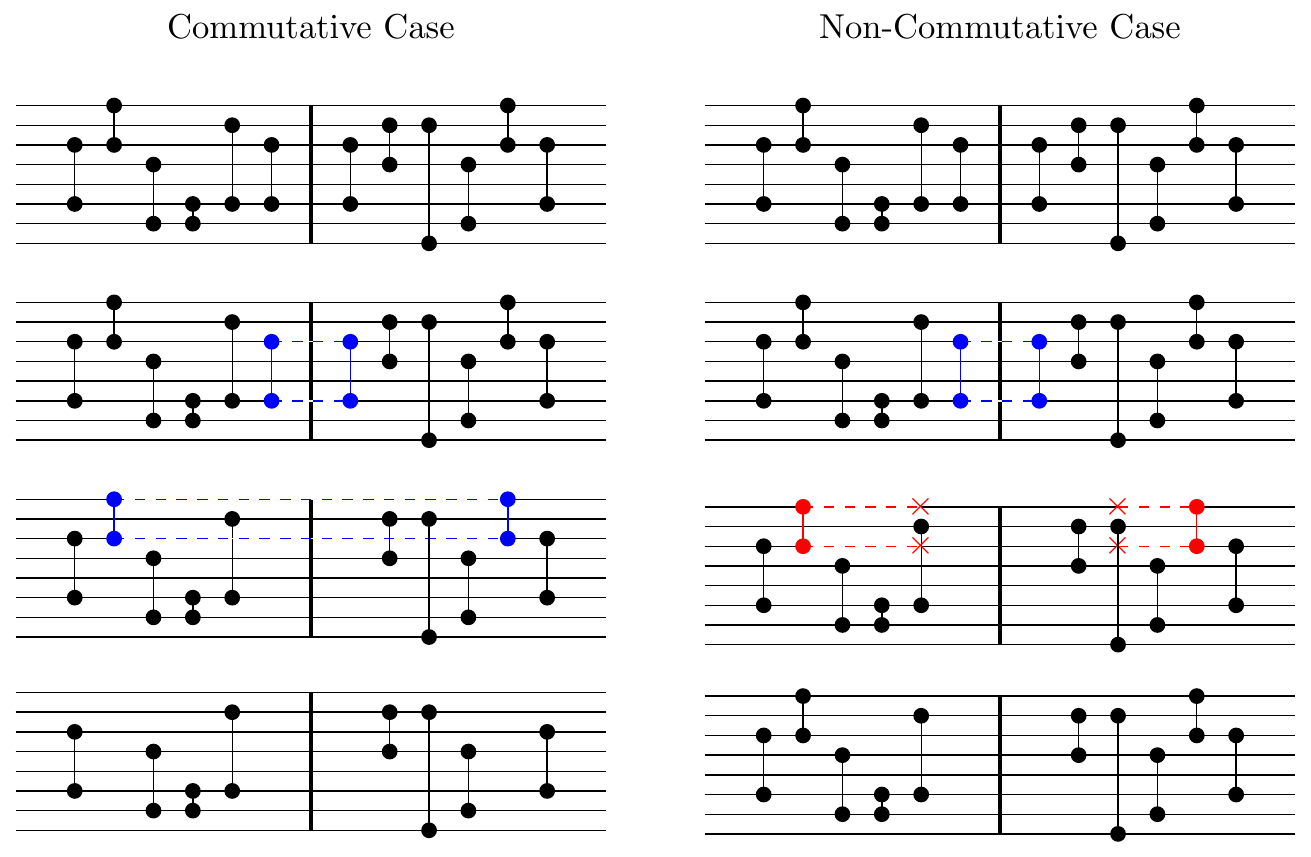}
    \end{center}
   \caption{This circuit represents the end of one copy of a circuit $Q_U$ implementing a hypothetical unitary $U$ and the beginning of a second copy of $Q_U$. In this plot horizontal lines are qubits, and the dots connected by vertical lines are gates acting on the pair of qubits they connect. For this illustration, we will consider gates to be their own inverse. Gates from two copies may cancel (illustrated here with dashed blue lines connecting the gates), reducing the complexity of the circuit and providing a more efficient way to compute $U^N$. This cancellation relies, however, on the ability of gates to commute past each other, so that gates which could cancel can meet. We argue that in the noncommutative case, fewer gates commute and so there are fewer cancelations of this type. In this illustration, we see on the third line that a gate which can commute to cancel in the commutative case is prevented from doing so in the non-commutative case due to mild non-locality. Cartoon inspired by one used in a talk by Adam Brown.} \label{fig:cartoon1}
\end{figure}

Now, what happens if we turn on non-commutativity in our theory? Let us suppose that our Hamiltonian $H = H_a$ varies continuously with the Moyal scale $a$, and suppose that our gates vary continuously as well so that the gates in the noncommutative theory can be identified with gates in the commutative theory. Suppose furthermore that for sufficiently small $\delta t$, $U_a(\delta t) = e^{iH_a \delta t}$ can be optimally approximated by the same circuit $Q$, but with each of the original gates $g$ replaced with its noncommutative analog $g_a$ (Call this circuit $Q_a$).
Then it is still true that $U_a^N$ can be implemented by $Q_a^N$. But now, because of the non-commutativity, it is likely that fewer of the gates at the beginning and end of $Q$ will commute with each other (see figure \ref{fig:cartoon1}). And so we can still write
\begin{equation}
\mathcal{C}_a(t)\approx \frac{t}{\delta t} \left[ \mathcal{C}_a(\delta t) - \chi_a \right] + \chi_a \approx  \frac{t}{\delta t} \left[ \mathcal{C}(\delta t) - \chi_a \right] + \chi_a ,
\end{equation}
but because fewer gates cancel, $\chi_a$ will be smaller than the original $\chi$. These mean that the complexifaction rate
\begin{equation}
\dot{\mathcal{C}}_a(t) \equiv \frac{d}{dt} \mathcal{C}_a(t)  \approx \frac{1}{\delta t} \left[ \mathcal{C}(\delta t) - \chi_a \right]
\end{equation} 
gets an enhancement due to the suppression of $\chi_a$. Finally we get an enhancement ratio of complexification rate as
\eq{\label{eq:ratio}
\dot{\mathcal{C}}_a(t) \approx \frac{\mathcal{C}(\delta t) - \chi_a}{\mathcal{C}(\delta t) - \chi} \dot{\mathcal{C}}(t).
}

The same effect could be understood as arising from an increased non-locality due to the noncommutativity.  The dependence of complexity growth on the locality of gates is explored in \cite{Brown:2015lvg}, where an extension of the Lloyd bound is studied by looking at the "$k$-locality" of the Hamiltonian and the gate set. A "$k$-local" operator is one that acts on at most $k$ degrees of freedom: a $k$-local Hamiltonian consists of interactions coupling at most $k$ degrees of freedom, and similarly a $k$-local gate set consists of at most $k$-local operators.
\footnote{To avoid dependence on the choice of basis, we would like to define $k$ as the maximum rank of the coupling terms, or the maximum rank of the generators of the gates.}
For convenience we let the Hamiltonian be "k-local" while the gate set is "j-local." Usually, the Lloyd bound should be satisfied if $j=k$, because one can choose the coupling terms as gates so that the time evolution could be easily implemented by the gates. However if one chooses a different $j$ for the gate set, a bound of the following general form is to be expected
\eq{
\dot{\mathcal{C}} \leq \frac{g(k)}{g(j)}\frac{2M}{\pi},
}
where $g(k)$ is a monotonically increasing function. The interesting connection to our interpretation of non-commutativity is that the Moyal area introduced in non-commutative space can be thought of as an effective $k$ for the Hamiltonian, meaning that non-local interactions couple wider range of degrees of freedom than local interactions. On the other hand, we are not changing $j$ because our holographic prescription is not changed. Then we have an extra factor $g(k)/g(j) > 1$ in the bound, hence an enhanced bound. A similar factor greater than 1 is hence obtained in eq(\ref{eq:ratio}).


\section{Finite Time behavior}
\label{sec:4}

Up to now, we have only discussed the asymptotic behavior of the complexification rate at late times. It is plausible that the early time complexification rate is not as important as the late time limit because there is a thermal scale time resolution for this quantity. One might think of this resolution as the time scale for a new gate to act on the state. In the paper \cite{Carmi:2017jqz} people carefully studied the finite time behavior of the complexification rate and found several interesting features. We will briefly outline the finite time behavior for noncommutative SYM, reproduce those features, and find new features introduced by the non-commutativity. 

We will rewrite equation (\ref{eq:finiteTimeRate}) using the dimensionless parameters
\begin{equation}
b = a r_H ,\quad \rho = r_B / r_H ,\quad \gamma = \frac{c \bar{c} \sqrt{\hat{g}_s}R^2}{\alpha'r_H^2},
\end{equation}
so that we get
\begin{equation}
\frac{\delta S}{\delta t} = \frac{\Omega_5 V_3 r_H^4}{(2\pi)^7 \hat{g}_s^2}
\bigg(\frac{-2\log(1+b^4 \rho^4)}{b^4}+4\rho^4+6+3(1-\rho^4)\log\big|\frac{\gamma \rho^2}{(1+b^4 \rho^4)^{1/4}(1-\rho^4)}\big|\bigg).
\end{equation} 
Note that since $T = r_H/\pi$, we have $b = \pi a T$. 

We will now normalize this by the late time commutative result at the same temperature to define
\begin{equation}\label{finitetime}
\dot{C}_\text{n}(\rho) = \frac{-\log(1+b^4 \rho^4)}{4 b^4} + \frac{1}{2}\rho^4 + \frac{3}{4} + \frac{3}{8}(1-\rho^4)\log\big|\frac{\gamma \rho^2}{(1+b^4 \rho^4)^{1/4}(1-\rho^4)}\big|
\end{equation}\label{eq:finitetime}

\begin{figure}[htbp]\label{FiniteTime1}
    \begin{center}
        \includegraphics[scale=0.28]{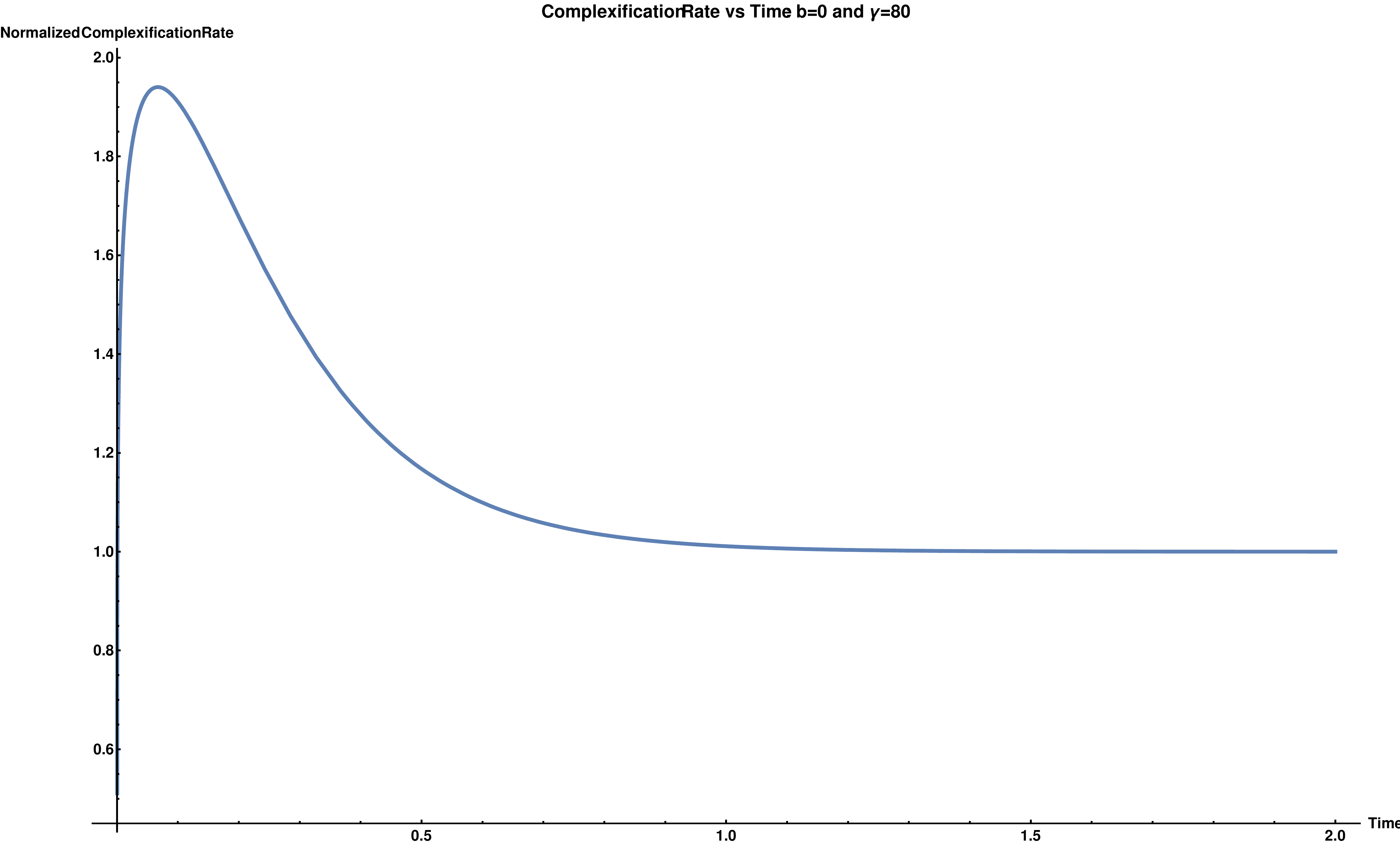}
    \end{center}
    \caption{Normalized complexification rate versus time in thermal units for $\gamma = 80$ and $b=0$.} \label{fig:FiniteTime1}
\end{figure}

Substituting left time in thermal units for $\rho$, can plot $\dot{C}_n$ vs time at fixed $b$ and $\gamma$, yeilding (in the case where we take $b\rightarrow 0$ and $\gamma = 80$) the plot in figure \ref{fig:FiniteTime1}. 

It is clear from this plot that there is a local maximum at early time (around $t = 0.1 \beta$, $\beta$ being the inverse temperature), and then at late times, it approaches the smaller asymptotic value from above. There is also a logarithmic divergence as $t$ goes to zero which comes from the log term in equation (\ref{finitetime}). Both of these features are observed in \cite{Carmi:2017jqz}, where they are discussed in great detail. The logarithmic divergence is not important in the sense that if you take the average complexification rate over a roughly thermal time scale, this divergence will be gone. A small period of decreasing complexity remains, but such behavior is not altogether prohibited. At early times the complexity is highly sensitive to the choice of the reference state, and only at late times is a constant growth rate expected for generic (time-independent) Hamiltonians. Regardless, the issues of the local maximum and the asymptotic approach to the "bound" from above are not resolved in any explanations here. One could average over an artificially long period of time to smooth out the local maximum, but doing so would never eliminate the approach from above, irrespective of the physicality of such a procedure.

\begin{figure}[htbp]\label{FiniteTime2}
    \begin{center}
        \includegraphics[scale=0.28]{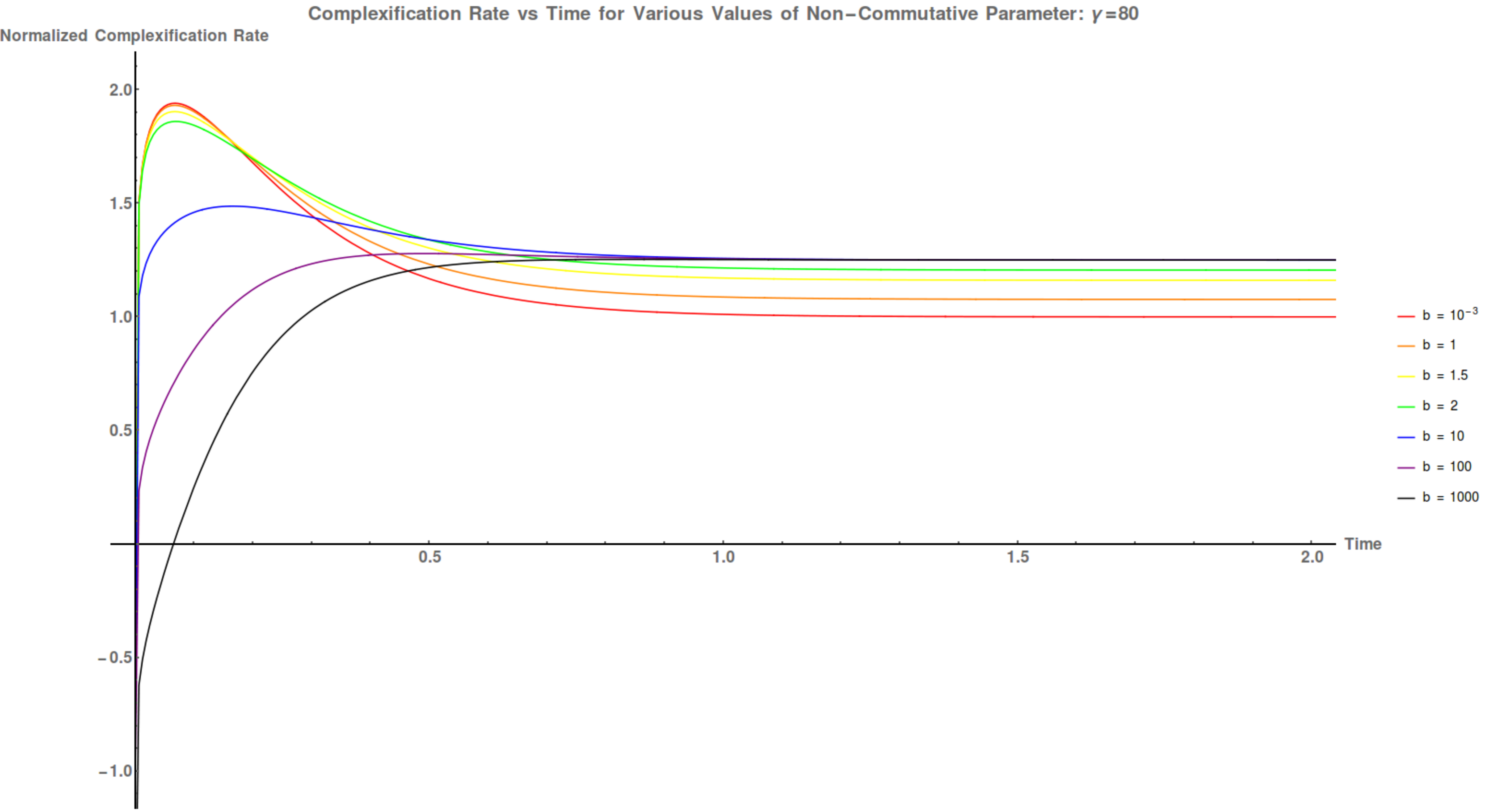}
    \end{center}
    \caption{normalized complexification rate versus time in thermal units. $\gamma$ is held fixed at 80 while $b=ar_H$ is varied.} \label{fig:FiniteTime2}
\end{figure}

Our primary interest here, however, is to discuss how these behaviors change with the noncommutative parameter $b$. To that end, we will consider what happens when we replot this curve fixing $\gamma$ but varying $b$. The result is displayed in figure \ref{fig:FiniteTime2}.

\begin{figure}[htbp]
    \begin{center}
        \includegraphics[scale=0.8]{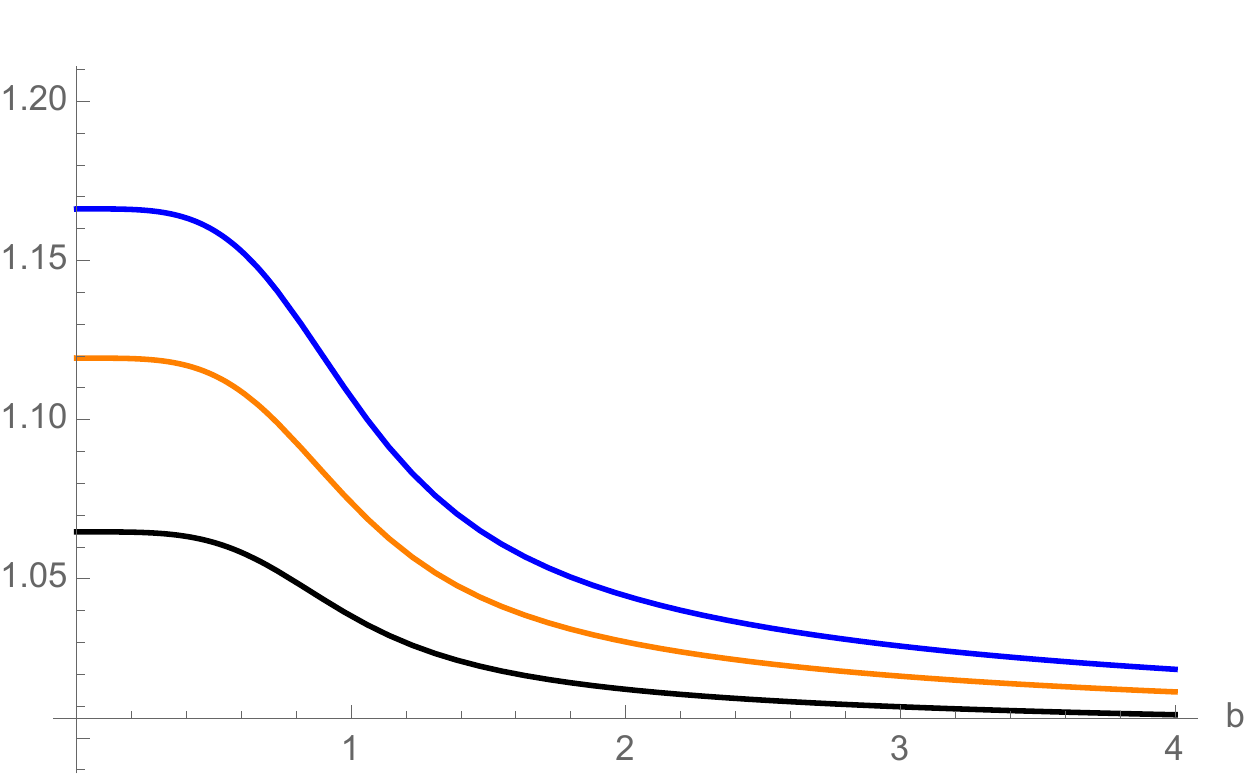} 
    \end{center}
    \caption{The vertical axis is the ratio between the local maximum and the asymptotic late time value of the complexification rate. The black, orange and blue curves correspond to $\gamma = 1,2,3$.} \label{fig:max_ratio}
\end{figure}

From figure \ref{fig:FiniteTime2} we see that as the non-commutativity is turned up, the local maximum decreases, and the asymptotic value increases. It is obvious that the change happens at $b\sim \pi$, which is when the Moyal scale $a$ is comparable to the thermal scale $T^{-1} = \pi/ r_H$. For $b\gg \pi$, it seems that the asymptotic value is approached from below. Strictly speaking, it is not true, because the local maximum always exists, but has a diminishing relative height and is pushed to very late time. We can find the local maximum and plot its ratio to the asymptotic value versus $b$ as in figure \ref{fig:max_ratio}. The fact that the local maximum decays physically rather than by tuning some artificial choice is a sign that the noncommutative complexification rate at late time is a more qualified bound for a generic quantum system. We will discuss it in more details in the conclusion.

\begin{figure}[htbp]\label{FiniteTime3}
    \begin{center}
        \includegraphics[scale=0.28]{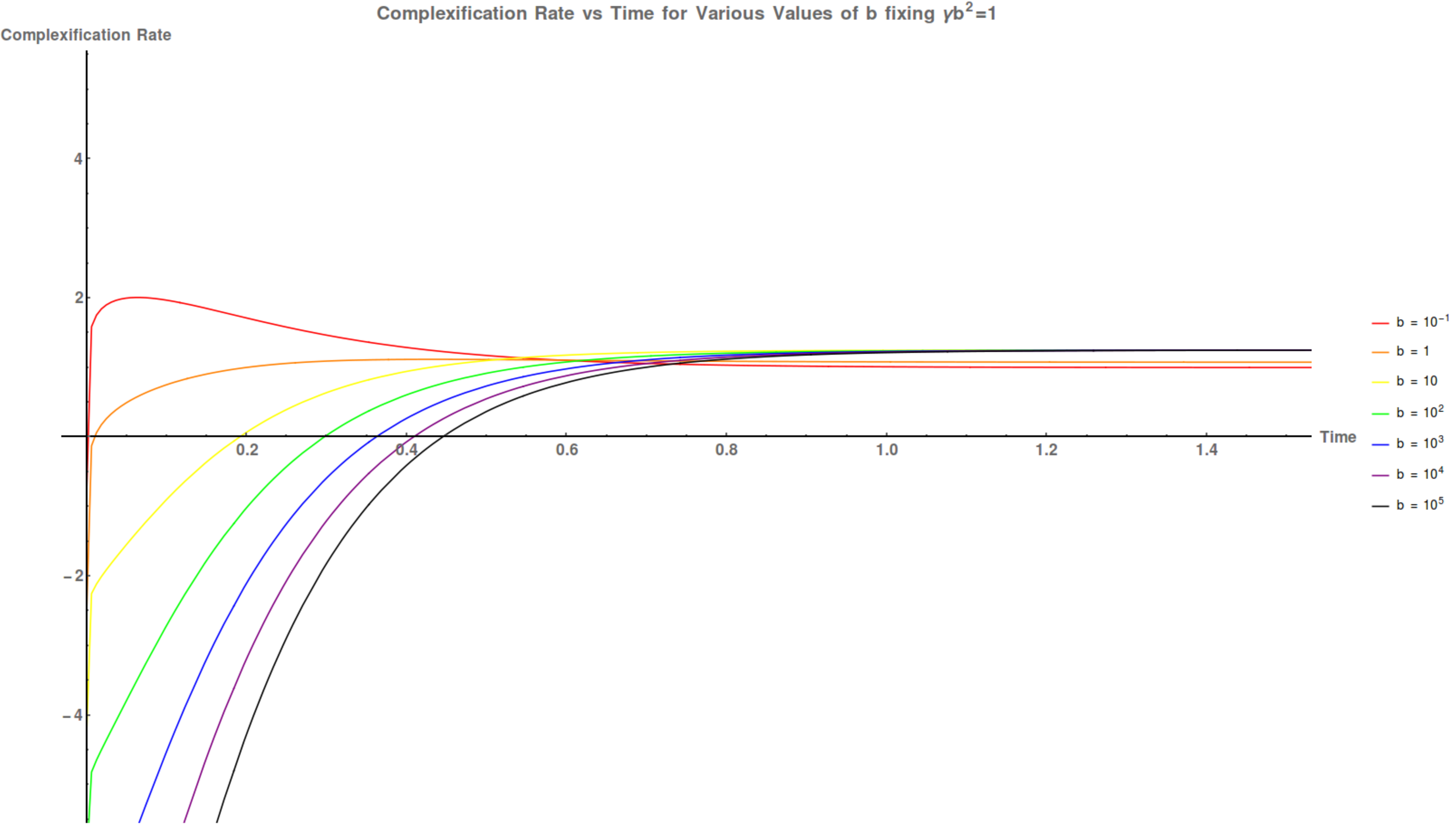}
    \end{center}
    \caption{normalized complexification rate versus time in thermal units. $\gamma b^2$ is held fixed at 1 while $b=ar_H$ is varied.} \label{fig:FiniteTime3}
\end{figure}

It is observed that the complexification rate mainly depends on temperature through the combination $b$, except an extra logarithmic contribution from $\gamma$. Therefore we expect that the variation with respect to temperature is similar to figure \ref{fig:FiniteTime2}. This can be implemented by varying $b$ while fixing the combination $\gamma b^2$, i.e., fixing $a$. When this is done with $\gamma b^2=1$ one gets figure \ref{fig:FiniteTime3}, which is indeed similar to figure \ref{fig:FiniteTime2}. This check shows that the only scale that the non-commutativity $a$ is sensitive to is the thermal scale. 

Finally, one may also be interested in the effect of $\gamma$, which at fixed AdS radius and temperature encodes information about the normalization of the generators of the null boundaries of the WDW patch. It has been suggested that this normalization, which is ambiguous in the action, should correspond to an ambiguity in the definition of complexity on the boundary such as the choice of reference state \cite{Jefferson:2017sdb}. In our case, we observe that the dependence on $\gamma$ does not depend on the non-commutativity at all, which seems to support this idea for a broader class of theories.


%


\section{Other noncommutative systems}
\label{sec:5}

As a test of the above argument, and to better understand the dependence of the enhancement on various factors, we would like to consider more examples of noncommutative field theories. 
It's easy to extend the D3 brane solution we discussed in Section \ref{sec:2} to other D$p$ branes, in which we are also able to put more noncommutative pairs of directions. For $p=4,5,6$, we can turn on more than one $B$ field component, making multiple pairs of directions non-commuting. Let us denote the number of non-vanishing $B$ components as $m$ so that $B$ will be a rank-$2m$ matrix. In this section, we will investigate the dependence of late time complexification rate on the dimension of space $p$ and the rank of the $B$ field.

\subsection{Supergravity solutions and decoupling limit}

The general string frame metric for non-extremal D$p$ branes with $m$ non-commuting pairs of directions are given as
\eq{
\frac{ds^2}{\alpha'} &= \left(\frac{r}{R}\right)^{\frac{7-p}2} \left(- f(r)dx_0^2 + \sum_{i=1}^{p-2m}dx_i^2 + \sum_{i=1}^m h_i(r)(dy_{i,1}^2+dy_{i,2}^2) \right) \\
& + \left(\frac{R}{r}\right)^{\frac{7-p}2} \left(\frac{dr^2}{f(r)} + r^2 d\Omega_{8-p}^2) \right)
}
where
\eqs{
f(r) = 1-\frac{r_H^{7-p}}{r^{7-p}}, \\
h_i(r) = \frac{1}{1+(a_ir)^{7-p}}.
}
In NS-NS sector we have
\eq{
&e^{2\Phi} = \hat{g}_s^2\left(\frac{R}{r}\right)^{\frac{(7-p)(3-p)}{2}}\prod_{i=1}^{m}h_i(r), \\
&B^{(i)} = -\frac{\alpha'}{(a_iR)^{\frac{7-p}{2}}}[1-h_i(r)] dy_{i,1} \wedge dy_{i,2}.
}
We also have many R-R fields turned on via the T-duality. One would expect them by looking at the Chern-Simons term in D brane action
\eq{
S_{CS}^{Dp} = \mu_p \int \left(C\wedge\exp(B+kF)\right)_{p+1}.
}
Only rank-$(p+1)$ R-R potential $C_{p+1}$ is turned on without any background field, whereas in the presence of $B$ field, terms like $C_{p+1-2n}\wedge B^{(i_1)}\wedge
\cdots\wedge B^{(i_n)}$ can also be sourced, where $n = 0,1,\cdots,m$. In other words, when $m=1$, we have $C_{p-1}$ turned on; when $m=2$, we have $C_{p-3}$ turned on, and so on. 

The general formulae for all these R-R fields are
\eq{
&C_{p+1} =                -\frac{(\alpha')^{\frac{p+1}{2}}}{\hat{g}_s}\left(\frac{r}{R}\right)^{7-p} \prod_{i}h_i(r) , \\
&C_{p-1}^{(j)} =         \frac{(\alpha')^{\frac{p-1}{2}}}{\hat{g}_s}\left(\frac{r}{R}\right)^{7-p} (a_jR)^{\frac{7-p}{2}}        \prod_{i\neq j}h_i(r), \\
&C_{p-3}^{(j,k)} =        -\frac{(\alpha')^{\frac{p-3}{2}}}{\hat{g}_s}\left(\frac{r}{R}\right)^{7-p} (a_ja_kR^2)^{\frac{7-p}{2}}     \prod_{i\neq j,k}h_i(r), \\
&C_{p-5}^{(j,k,l)} =    \frac{(\alpha')^{\frac{p-3}{2}}}{\hat{g}_s}\left(\frac{r}{R}\right)^{7-p} (a_ja_ka_lR^3)^{\frac{7-p}{2}}\prod_{i\neq j,k,l}h_i(r).
}
We are omitting the basis here, but it's clear that these components are along all the directions on D$p$ brane except for the directions of the $B$ fields indicated by their superscript. We also omitted their (inverse) hodge dual forms which may contribute to the action. 

While these are all good solutions for supergravity in the bulk, one has to be careful with its world volume dual theory. The decoupling limit of the world volume theories for $2\leq p\leq 6$ in the presence of $B$ field is studied in \cite{Alishahiha:2000qf}, with the conclusion that there is no decoupling limit for D6 branes even for $m>0$. For $p\leq 5$, decoupling limits do exist, and it's reasonable to talk about the complexity on the world volume theory. One may be worried that for D4 brane we have to up lift to 11 dimensions to compute the M theory action, but the effective string coupling at high energy is
\eq{
e^{\Phi} \sim r^{\frac{(7-p)(p-3-2m)}{4}},
}
which is suppressed by the non-commutativity when $m\geq 1$, indicating that at sufficiently high energy, we don't have to go to M theory. 

As such, we will be using type IIB action for odd $p$ and type IIA action for even $p$. The type IIA action is
\eq{
S_{\rm string}^{\rm IIA} = \frac{1}{2\kappa^2}\int dx^{10} \sqrt{-g} \left[e^{2\Phi}(\mathcal{R} + 4|d\Phi|^2 - \frac{1}{2}|H|^2) -\frac{1}{2}|F_2|^2 -\frac{1}{2}|\tilde{F}_4|^2 \right] -\frac{1}{4\kappa^2}\int B\wedge F_4\wedge F_4,
}
with the usual conventions:
\eq{
F_2 = dC_1,\qquad F_4 = dC_3,\qquad \tilde{F}_4 = F_4 - C_1\wedge H.
}

\subsection{Complexification Rates}

We report the action growth rates with the following $p$-dependent prefactor,
\eq{\label{eq:cp}
    c_p \equiv \frac{\Omega_{8-p} r_H^{7-p}}{(2\pi)^7\hat{g}_s^2},
}
We also divide out the transverse volume $V_p$ to give a "density of action." The complexification rate will be related to the action growth rate by eq(\ref{eq:prop_k}), where the coefficient $k$ is not specified yet. We will discuss the strategy of choosing $k$ at the end of the section. Both the joint and boundary contributions to the late time complexification rate take a particularly simple form:

\begin{equation}\begin{split}
        \dot{S}_{\text{joint}}&=(7-p)c_p\\
        \dot{S}_{\text{boundary}}&=\frac{1}{8}(65-14 p+p^2)c_p\\
\end{split}\end{equation}

The bulk contributions exhibit more interesting dependencies on the size and number of noncommutativity parameters. These are here reported for each $p$.

\subsubsection*{D2 Brane}
This is the simplest case, where we have fewest R-R fields and don't need to put the self-duality constraint. We have
\eqs{
    &F_2 = dC_{p-1}, \\
    &\tilde{F}_4 = dC_{p+1} - C_{p-1}\wedge H.
}

Plugging them in the type IIA action, we obtain the complexity growth rate. Including all contributions, the late time limit becomes
\eq{
    \dot{S}_{p=2,m=1} = 12 c_p.
}
Surprisingly, we find that the late time complexification rate does not even depend on the non-commutativity parameter $a$. We may argue that it is the case where the bound is already saturated so that non-commutativity could not enhance it anymore.

\subsubsection*{D4 Brane}
This is the minimal dimension that we can include two pairs of noncommutative directions, hence $m=2$. The R-R field contents are
\eqs{
    &F_2 = dC_{p-3}^{(1,2)}, \\
    &\tilde{F}_4 = \sum_i\left[dC_{p-1}^{(i)} - C_{p-3}^{(1,2)}\wedge H^{(i+1)}\right] + *^{-1}\left[dC_{p+1} - \sum_i\left(C_{p-1}^{(i)}\wedge H^{(i)}\right)\right].
}
Note that mod $m$ is understood in the supercript of the forms.

The complexity growth rate including all contributions has late time limit
\eq{
    \dot{S}_{4,2} = \left(5+\frac{3 a_1^3 a_2^3 r_H^6}{(1+a_1^3 r_H^3)(1+a_2^3 r_H^3)}\right)c_p.
}

The $p=4,\ m=0,1$ cases can be obtained by taking one or both of the $a$ parameters to zero:
\eq{
    \dot{S}_{4,0}=\dot{S}_{4,1} =5 c_4
}
It's striking that turning on a single pair of noncommutative directions does not affect the late time complexification rate at all, but turning on the second pair does increase the rate. It means that we cannot use the argument as for $p=2$ to explain the zero enhancement here because obviously the bound was not saturated yet.

\subsubsection*{D5 Brane}
It's another case where we need to take into account the self-duality issue. Again we can have $m=2$, and the R-R field contents are
\eqs{
&F_3 = dC_{p-3}^{(1,2)} + *^{-1}\left[dC_{p+1} - \sum_i\left(C_{p-1}^{(i)}\wedge H^{(i)}\right)\right], \\
&\tilde{F}_5 = \sum_i\left[dC_{p-1}^{(i)} - \frac{1}{2}C_{p-3}^{(1,2)}\wedge H^{(i+1)} + \frac{1}{2} dC_{p-3}^{(1,2)}\wedge B^{(i+1)}\right] + \text{self dual}.
}

The complexity growth rate including all contributions has late time limit
\eq{
\dot{S}_{5,2} = 
\left(\frac{11}{2}+\frac{a_1^2a_2^2r_H^4-2}{2(1+a_1^2r_H^2)(1+a_2^2r_H^2)}+\frac{a_2^2\log(1+a_1^2r_H^2)}{2a_1^2(a_1^2-a_2^2)}+\frac{a_1^2\log(1+a_2^2r_H^2)}{2a_2^2(a_2^2-a_1^2)}\right)c_5.
}
We can also examine the $m=1$ case by taking $a_2 = 0$ and $a_1 = a$:
\eq{
\dot{S}_{5,1} = \left(5 - \frac{1}{1+a^2 r_H^2} \right)c_p, \\
\dot{S}_{5,0} = 4c_p
}
In contrast with $p=4$, turning on the first pair of noncommutative directions already changes the complexity, and turning on the second enhances more.

\subsubsection*{D6 Brane}
Finally we may investigate a case where we can turn on 3 pairs of noncommutative directions, hence D6 brane. For $m=3$, the R-R field contents are
\eqs{
&F_2 = dC_{p-5}^{(1,2,3)} + *^{-1}\left[dC_{p+1} - \sum_i\left(C_{p-1}^{(i)}\wedge H^{(i)}\right)\right], \\
&F_4 = \sum_i\left[dC_{p-3}^{(i+1,i+2)} - C_{p-5}^{(1,2,3)}\wedge H^{(i)}\right] + *^{-1}\left[dC_{p-1}^{(i)} - \sum_{j\neq i}C_{p-3}^{(i,j)}\wedge H^{(j)}\right].
}

The complexity growth rate including all contributions has late time limit
\eq{
\dot{S}_{6,3} =\left(4 
+\frac{a_1a_2\log(1+a_3r_H)}{(a_2-a_3)a_3(a_3-a_1)r_H}
+\frac{a_2a_3\log(1+a_1r_H)}{(a_3-a_1)a_1(a_1-a_2)r_H}
+\frac{a_3a_1\log(1+a_2r_H)}{(a_1-a_2)a_2(a_2-a_3)r_H}\right)
c_6
}
The three $a$-dependent terms have the property that no matter how many $a$'s you turn off, their sum is a constant as -1. Thus again, it is a situation where only turning on maximum number of non-commutativity can we increase the non-commutativity, similar to the $p=4$ case. 

\eq{\dot{S}_{6,0}
    =\dot{S}_{6,1}
    =\dot{S}_{6,2} = 3 c_6
}

However, this complexity growth rate seems to have no physical meaning, because there is not a world volume theory that is decoupled from gravity. The holographic principle is subtle in this case. We present the result here because the bulk computation can be done in a similar manner without noting the difference. Whether the quantity so computed has any physical meaning is an open question.
    
\subsection{Summary of Results}

From the above computation, we find that when we turn on non-commutativity on Dp branes, the complexity growth rate either stays the same, or increases. The fact that it does not decrease is encouraging for our argument given in section \ref{sec:3}. However, the values of the enhancement ratio are not understood.

In the table \ref{tab:general_p}, we list all the density of late time action growth rate in unit of $c_p$, in the limit that all $m$ non-commutativity parameters $a_i$, $i=1,..,m$, goes to infinity.
\begin{table}
    \centering
    \begin{tabular}{l | c  c  c  c | c}
        $p$ & $m=0$ & $m=1$ & $m=2$ & $m=3$ & $\pi B_L$ \\
        \hline
        2 & 12 & 12 & - & - & 7\\
        3 & 8 &10 & - & - & 6\\
        4 & 5 & 5 & 8 & - & 5\\
        5 & 4 & 5 & 6 & - & 4\\
        6 & 3 & 3 & 3 & 4 & 3
    \end{tabular}
    \caption{This table lists all the action growth rate at late time for general $p$ and $m$. They are in unit of the constant $c_p$ defined in eq(\ref{eq:cp}). The last column is showing the Lloyd bound $B_L$ also in unit of $c_p$.}\label{tab:general_p}
\end{table}

There are no obvious laws that govern these rates in general, but we do observe some interesting features. For both D3 and D5 branes, we have enhancement from each pair of non-commuting directions. In particular, the ratio for the enhancement from the first pair are the same in both cases, and the enhanced amount from the first and second pair are also the same in D5 brane. These two cases seem to provide reasonable behaviors one may naively expect. On the other hand, the type IIA supergravity with even-$p$s does not always have complexification rate enhancement from non-commutativity. The reason for it may depend on the details of the boundary theory.

In the table \ref{tab:general_p}, we also list the Lloyd bound computed from the ADM mass of the geometry (see Appendix \ref{app:3}). 
One may set the coefficient $k$ in eq(\ref{eq:prop_k}) to let any of the complexification rate to saturate the Lloyd bound. For instance, if we want to set the commutative $\mathcal{N} = 4$ SYM ($p=3$, $m=0$) to saturate the bound, we can take $\pi k_{p=3} = 3/4$. However, the consequence is that we can always turn on the non-commutativity and violate this bound. In order that the Lloyd bound is not violated, we need to guarantee that the maximum complexification rate for each $p$ is bounded by $B_L$, thus
\eq{
k_2 \leq \frac{7}{12\pi}, \quad k_3 \leq \frac{3}{4\pi}, \quad k_4 \leq \frac{5}{8\pi}, \quad k_5 \leq \frac{2}{3\pi}, \quad k_6 \leq \frac{3}{4\pi}.
}

If one follows the argument at the end of section \ref{sec:3}, and get an enhanced bound for non-commutative field theory, the bound on $k_p$ will be weaker. On the other hand, the Lloyd bound is defined under the assumption that all gates take a generic state to an orthogonal state, which is usually not true. It is argued that we simply should not take this bound seriously \cite{Cottrell:2017ayj}. This objection will make it hard to determine what $k$ should be, but for our purpose, $k$ does not affect our main results.


\section{Conclusion}
\label{sec:Conclusion}

In this paper, we have considered the effects of non-commutativity on the holographic complexity of SYM according to the complexity = action conjecture. We have done this in the hope that this would produce further evidence about the validity of this conjecture, and of the concept of holographic complexity more generally. Our main result is that the late time complexification rate increases with the non-commutativity in a class of theories. 


We computed the holographic complexity for 4D $\mathcal{N}=4$ non-commutative super Yang-Mills, by evaluating the WDW action in the bulk geometry described by type IIB supergravity with D3 branes. We saw a 5/4 enhancement for late time complexification rate in the non-commutative result over the commutative result. 
This was striking because it is well known that the thermodynamics of this theory are independent of the non-commutative parameter $a$. The observed changes to complexity support the idea that complexity is more than thermodynamics, and indicates that the CA prescription is reproducing this feature of complexity.
Comparing to the Lloyd bound derived from the total energy, we discovered that using the coefficient of proportionality $k = 1/\pi$ as in \cite{Brown:2015bva} will make the commutative late time complexification rate violate the bound. One could in principle avoid this by arguing that $k$ should not be universal for all kinds of theories, but the commutative black hole still can not saturate the bound because there should be space for enhancement from the non-commutativity.

We presented a quantum argument to explain this enhancement and to argue that we should have expected it. We assume that the time evolution operator is approximated by sequential copies of the same quantum circuit, and the optimization of the total circuit when you combine them will be less efficient in non-commutative theories. 
We also argue that this expectation matches the $k$-locality model prediction if we relate the size of Moyal scale to the size of locality $k$.
Then we investigate the finite time behavior of this complexification rate and see that the problematic finite time maximum gets suppressed by non-commutativity. 

Finally, we generalized the solution for D3 branes to general D$p$ branes to get a broader class of noncommutative gauge theories. We presented similar calculations as for $p=3$ and obtained the late time complexification rates for $2 \leq p \leq 6$ and all allowed ranks of the $B$ field. 
The results for $p=5$ are similar to those for $p=3$ but can have another enhancement of the same magnitude from a second $B$ field component.  This is consistent with our heuristic argument. The results for the even $p$ cases are less well understood.
We found that there is no enhancement for $p=2$ and that for $p=4$ one must introduce a second $B$ field component to get an enhancement. This result would seem to be in mild tension with the argument of section \ref{sec:3}. 
%
The correct explanation for this behavior is left for future work. Despite not seeing an enhancement in some cases, it is at least encouraging that no decrease was observed, which would have been a much clearer contradiction to the arguments of section \ref{sec:3}.

Regarding the statement that non-commutativity enhances the complexification rate in general, there are several interesting aspects one can investigate. First, this result is in tension with the often expressed idea that the commutative AdS-Schwarzschild black hole is the fastest possible computer \cite{Brown:2015bva}. If non-commutativity can somehow increase the computational speed even further, it would be very interesting to see if it also increases the scrambling process of the black hole. 
%
Second, it also would be interesting to compute the complexity of a weakly coupled field theory on a non-commutative manifold in order to test the conclusion of our heuristic argument in a non-holographic context. Such a computation would, in light of this work, provide for a more robust check on the complexity = action conjecture. The work of \cite{Jefferson:2017sdb, Hashimoto:2017fga} might prove useful to such an endeavor. 
%
%
Another interesting extension of this work would be to repeat the computations for the complexity = volume, and the complexity = spacetime volume conjectures, which will be both a test for our results and a test for the holographic complexity prescriptions.
Finally, it was pointed out to us by Eoin \'O Colg\'ain that the geometry corresponding to the $D3$-brane case that we have considered here has been discovered to belong to a larger class of deformations of $AdS_5$, studied in e.g. \cite{Araujo:2017jkb, Araujo:2017jap, Bakhmatov:2017joy}. It would perhaps be interesting to extend the results of this paper to the more general case.


\section*{Acknowledgments}
We would like to thank Jacques Distler, Hugo Marrochio, Robert Myers, and Phuc Nguyen for useful comments and discussion. This material is based upon work supported by the National Science Foundation under Grant Number PHY-1620610.

\appendix

\section{Calculation of $\dot{S}_{WDW}$}
\label{app:1}
To minimize clutter in expressions, in this appendix we set $2\kappa^2 = (2\pi)^7 \alpha^4 =1$ and reinstate $\kappa$ dependence only at the end. Following the systematic treatment of \cite{Lehner:2016vdi}, the action on a bulk subregion is divided into contributions as follows:

\begin{equation}\begin{split}\label{eq:GeneralAction}
S_{\mathcal{V}}=&\int_\mathcal{V}\big(\mathcal{R} +\mathcal{L}_m\big)\sqrt{-g}dV \\
&+2 \Sigma_{T_i}\int_{ \partial\mathcal{V}_{T_i}} K d\Sigma
+2 \Sigma_{S_i}\text{sign}(S_i)\int_{ \partial\mathcal{V}_{S_i}} K d\Sigma
-2 \Sigma_{N_i}\text{sign}(N_i)\int_{ \partial\mathcal{V}_{N_i}} \kappa d\sigma d\lambda\\
&+2 \Sigma_{j_i}\text{sign}(j_i)\int_{B_{j_i}} \eta_{j_i} d\sigma
+2 \Sigma_{m_i}\text{sign}(m_i)\int_{B_{m_i}} a_{m_i} d\sigma
\end{split}\end{equation}

The first line we call the bulk contribution.  The second line contains boundary contributions along timelike ($\mathcal{T}$), spacelike ($\mathcal{S}$), and null boundaries ($\mathcal{N}$), respectively.  The final line contains joint contributions, divided into those which result from intersections of timelike and/or spacelike boundaries, and those which include one or more null boundaries. Sign conventions and notation for integrand quantities will be explained as needed in what follows.

While the action on a WDW patch is obviously of interest for its conjectured relation to Quantum Complexity, its time derivative is simpler to compute and interesting for diagnostic purposes. Due to the spacetime symmetries, this quantity reduces to the difference of two volume contributions ($\mathcal{V}_1$ and $\mathcal{V}_2$ in figure \ref{fig:WDW}), one boundary surface contribution ($\mathcal{S}_\epsilon$ in figure \ref{fig:WDW}), and two joint contributions ($B_1$ and $B_2$ in figure \ref{fig:WDW}).

\begin{equation}\begin{split}\label{eq:dSdt}
&\delta S_{\rm WDW} = \delta S_{\text{bulk}} +\delta S_{\text{boundary}} +\delta S_{\text{joints}}\\
\vspace{2mm}
&\delta S_{\text{bulk}}=S_{\mathcal{V}_1}-S_{\mathcal{V}_2}\\
&\delta S_{\text{boundary}}=-2 \int_{\mathcal{S}_\epsilon} K d\Sigma\\
&\delta S_{\text{joints}}=2 \int_{B_1}a_1 d\sigma -2 \int_{B_2}a_2 d\sigma\\
\end{split}\end{equation}

\subsection{Bulk Contribution}
The bulk integral contributions are of the form:
\begin{equation}
S_{\rm bulk}=\int_\mathcal{V}\sqrt{-g_E}\big(\mathcal{R} + \mathcal{L}_m\big)d\mathcal{V},
\end{equation}
where Einstein frame metric is used. For the action eq(\ref{eq:IIB-action}) and field content eq(\ref{eq:field solutions}) we have
\eqs{
    &\mathcal{R}=\frac{-2\sqrt{\hat{g}_s}\big(2a^4r_H^4+a^8r^4(r^4+r_H^4)\big)}{\alpha' R^2(1+a^4r^4)^{9/4}}, \\
    &\mathcal{L}_m=\frac{2\sqrt{\hat{g}_s}\big(4a^4r_H^4+a^8r^4(3r^4+r_H^4)\big)}{ \alpha' R^2(1+a^4r^4)^{9/4}}.
}

We let the integral over $x_1$, $x_2$, and $x_3$ give $V_3$ and the five-sphere $\Omega_5$. Also abbreviate  $C=\dfrac{\alpha^{\prime 4} \Omega_5 V_3}{\hat{g}_s^2}$.  Further let $\rho(u,v)$ and $\bar{\rho}(u,v)$ denote the radial value $r$ as implicit functions of advanced/retarded coordinates $u$ and $v$ from the appropriate quadrant (here the left and bottom quadrants, respectively). The form of these functions is not important here. 

The bulk contribution for $\mathcal{V}_1$ can be written in $(u,r)$ coordinates with radial limits expressed implicitly. 

\eq{
    S_{\mathcal{V}_1} = 
    C\int_{u_L}^{u_L+\delta t}du 
    \int_{r=\epsilon}^{r=\rho_L(u-v_L)}dr
    \frac{4r^3(a^8r^8+a^4 r_H^4)}{(1+a^4 r^4)^2} , 
}
Here $r=\epsilon$ is a surface close to the singularity which will be sent to zero. A similar expression can be written for $\mathcal{V}_2$ in $(v,r)$ coordinates, and after the radial integration we have:  

\begin{equation}
\frac{1}{C}(S_{\mathcal{V}_1}-S_{\mathcal{V}_2})=
\int_{u_L}^{u_L+\delta t}du\big(G(\rho_L(u-(v_L+\delta t)))-G(\epsilon)\big)
-\int_{v_L}^{v_L+\delta t}dv\big(G(\rho_L(u_L-v)))-G(\bar{\rho}(u_R,v))\big)
\end{equation}

Changing variables $u\rightarrow u_L+v_L-v+\delta t$ leads to a cancellation of terms such that for small $\delta t$ we are left with
\begin{equation}\begin{split}
S_{\mathcal{V}_1}-S_{\mathcal{V}_2}&\approx C \bigg(G\big(\bar{\rho}(u_R,v_L)=r_B\big)-G\big(\epsilon\big)\bigg)\delta t,\\
G(r)&=\frac{a^4(2r^4+a^4L^8-r_H^4)-2(1+a^4r^4)\log(1+a^4r^4)}{(a^4+a^8r^4)}.
\end{split}\end{equation}
This cancellation is expected from the boost symmetry of the left wedge of the spacetime, and also indicates the cutoff independence of our calculation.
We denote the radial value at the bottom corner of the WDW patch $\bar{\rho}(u_R,v_L) \equiv r_B$. As $\epsilon\rightarrow 0$ we find a bulk contribution of
\begin{equation}\label{eq:p3bulk}
\dot{S}_{\rm bulk}=\lim_{\delta t\to 0}\frac{S_{\mathcal{V}_1}-S_{\mathcal{V}_2}}{\delta t} = \frac{\alpha^4 \Omega_5 V_3}{\hat{g}_s^2} \bigg(\frac{a^4r_B^4}{1+a^4r_B^4}(r_H^4-r_B^4) + 2r_B^4 - \frac{2\log(1+a^4r_B^4)}{a^4}\bigg)
\end{equation}
Note that $r_B$ is related to $t_L$ in the manner that as $t_L\to\infty$, $r_B\to r_H$. Therefore, the late time limit can be obtained by taking $r_B\to r_H$ limit.

\subsection{Boundary Contributions}

We adopt the convention that the null boundary geodesics are affinely parameterized: $k^\mu \nabla_\mu k^\nu = \kappa k^\nu$ with $\kappa=0$, which simplifies the action computation considerably because all but one boundary surface ($\mathcal{S}_\epsilon$) make no contribution. The boundary $\mathcal{S}_\epsilon$ is the spacelike surface $r=\epsilon\rightarrow 0$. The contribution is of the form

\begin{equation}\begin{split}
\delta S_\mathcal{\text{boundary}} =-2\int_{ \mathcal{S}_\epsilon} K d\Sigma
\end{split}\end{equation}
where $d\Sigma$ is the induced volume element on the boundary hypersurface and $K$ is the extrinsic curvature: $K = g^{\mu \nu}\nabla_\mu s_\nu $ with the unit normal $s^\nu$ chosen to be future directed, away from the WDW patch.  This convention for choosing the direction of the surface normal is responsible for the minus sign on this term \cite{Lehner:2016vdi}.

For our metric eq(\ref{metric_E}) we have 
\begin{equation}
K=\left(\frac{\hat{g}_s}{\alpha^2}\right)^{1/4}
\frac{4r h(r) f'(r)+f(r)\left(32 h(r)-r h'(r)\right)}{8 R h(r)^{7/8}\sqrt{-f(r)}},
\end{equation}
which as $\epsilon\rightarrow 0$ leads to 
\begin{equation}\label{eq:p3spacelike}
\dot{S}_\mathcal{\text{boundary}} = 4 r_H^4 \frac{\alpha^4 \Omega_5 V_3}{\hat{g}_s^2}
\end{equation}

\subsection{Joint Contributions}\label{appsubsec:joints}
There are two joints ($B$ and $B'$) which contribute to the complexification rate. Each of these is comprised of the intersection of two null surfaces, so their contributions are of the form

\begin{equation}\label{eq:joint}
S_{J}=2\Sigma_{m_i}\text{sign}(m_i)\int_{B_{m_i}} a_{m_i} d\sigma
\end{equation}

\[ a_{m_i} = \log \left| -\frac{1}{2}k_L\cdot k_R \right| \]
where $dS$ is the volume element on the joint.
Here $k_L$ and $k_R$ are future-pointing null generators along the left-moving and right-moving boundaries, respectively. Both of the joints in question lie at the past of the corresponding null segments, which together form the past boundary of a WDW patch.  Together these facts determine that the sign of each joint's contribution to the WDW patch action is positive \cite{Lehner:2016vdi}, and so taking a difference of two patches leads to the signs given in equation \ref{eq:dSdt}.

In addition to the affine parameterization of boundary generators, a convention must be chosen to fix their normalization. It may be possible to associate the freedom allowed by this choice with corresponding conventions which must be established in the definition of quantum complexity (e.g., choice of reference state and gate set). Indeed, progress has been made in this direction \cite{Jefferson:2017sdb}.  For our purposes, establishing a normalization convention is necessary to make meaningful comparisons between different WDW patch actions (such as that implicit in our ``time derivative") as parameters of the theory are adjusted.

We normalize according to $k_L \cdot t_L =-c$ and $k_R \cdot t_R=-\bar{c}$, where $\hat{t}_R$ and $\hat{t}_L$ are normalized generators of time-translation on each boundary.  With this in mind we choose

\begin{equation}\begin{split}\label{eq:nullnormalization}
(k_L)_\mu&=-c(\delta^t_\mu-\sqrt{\frac{-g_{rr}}{g_{tt}}}\delta^r_\mu)\\
(k_R)_\mu&=\bar{c}(\delta^t_\mu+\sqrt{\frac{-g_{rr}}{g_{tt}}}\delta^r_\mu).
\end{split}\end{equation}

For small $\delta t$, the joints $B_2$ and $B_1$ are at fixed radii $r= r_B$ and $r=r_B +\frac{1}{2}\sqrt{\frac{-g_{tt}}{g_{rr}}}\delta t $, respectively. The quantities $a_m$ in equation \ref{eq:joint} are easily evaluated at each joint and the combined contribution is found to be:

\begin{equation}\label{eq:p3joints}
\begin{split}
S_{B_1}-S_{B_2}&=2 \frac{\alpha^4 V_3 \Omega_5}{\hat{g}_s^2}\bigg(r^3\log\big[-\bar{c}c \frac{R^2(\hat{g}_s^2 h(r))^{1/4}}{\alpha r^2 f(r)}\big]\bigg|^{r=r_{B_1}}_{r=r_{B_2}}\bigg)\\
&\approx \delta t\frac{\alpha^4 V_3 \Omega_5}{\hat{g}_s^2} \bigg(\frac{2r_H^4+r_B^4(2+a^4(3r_B^4+r_H^4))}{1+a^4r_B^4}+3(r_H^4-r_B^4)\log\bigg|\frac{c \bar{c} \sqrt{\hat{g}_s}R^2 r_B^2}{\alpha (1+a^4r_B^4)^{1/4}(r_H^4-r_B^4)}\bigg| \bigg)
\end{split}
\end{equation}

\subsection{Combined Contributions}

We can combine contributions \ref{eq:p3bulk}, \ref{eq:p3spacelike}, and \ref{eq:p3joints} to arrive at the full time rate change of the WDW patch action (reinstating explicit $\kappa$ dependence and immediately using $2\kappa^2=(2\pi)^7\alpha^4$):
\begin{equation}\begin{split}
\dot{S}_{WDW}&=\frac{\Omega_5 V_3}{(2\pi)^7\hat{g}_s^2}
\bigg(\frac{-2\log(1+a^4r_B^4)}{a^4}+4r_B^4+6r_H^4+3(r_H^4-r_B^4)\log\big|\frac{c \bar{c}\sqrt{\hat{g}_s}R^2 r_B^2}{\alpha (1+a^4 r_B^4)^{1/4}(r_H^4-r_B^4)}\big|\bigg)
\end{split}\end{equation} 

%
%
%
%

\section{Thermodynamics and the Lloyd Bound}
\label{app:3}
It is interesting that the thermodynamic quantities for these systems exhibit no dependence on the noncommutativity parameter $a$ (see \cite{Maldacena:1999mh} for discussion).  We find that for general $p$\footnote{Note that for $p=5$ equations \ref{eq:thermo} would indicate zero specific heat.  We take this as further evidence that results for $p\ge5$ should be viewed skeptically.}
\begin{equation}\begin{split}\label{eq:thermo}
E&=\frac{(9-p)r_H^{(7-p)}}{2 (2\pi)^7 \hat{g}_s^2}V_p\Omega_{(8-p)}\\
T&=\frac{(7-p)r_H^{(5-p)/2}}{4 \pi R^{(7-p)/2}}\\
S&=\frac{4\pi R^{(7-p)/2} r_H^{(9-p)/2}}{(2\pi)^7 \hat{g}_s^2}V_p\Omega_{(8-p)}
\end{split}\end{equation}
with $E$ being the ADM mass.  The first law $dE=TdS$ is easily confirmed.

In the original CA duality conjecture \cite{Brown:2015bva,Brown:2015lvg} the proportionality constant in $\text{Complexity} =k \times \text{Action}$ was fixed by an expectation that black holes are the fastest computers in nature, and that at late times they would saturate a bound from Lloyd \cite{Lloyd, Margolus:1997ih}.  Matching $\dot{C}=\frac{2 M}{\pi}$ at late times for Schwarzschild AdS black holes sets the constant at $k=\frac{1}{\pi}$. The relevance of the Lloyd bound to these considerations is questionable \cite{Cottrell:2017ayj}, but in the interest of comparison we note that the systems studied in this work would require different constants to meet the same criterion: for the commutative black holes to saturate the bound at late times, $k=\underset{t\rightarrow \infty}{\lim} \frac{2M}{\pi \dot{S}}$ would be given by

\begin{center}
\begin{tabular}{ |c|c|c|c|c|c| }
    \hline
      & $p=2$  & $p=3$  &$p=4$  &$p=5$  &$p=6$\\
    \hline
    $k$ &$\frac{7}{12\pi}$ &$\frac{3}{4\pi}$  &$\frac{1}{\pi}$  &$\frac{1}{\pi}$  &$\frac{1}{\pi}$\\
    \hline
\end{tabular}
\end{center}

Furthermore, if the proportionality $k$ were fixed with reference to commutative black holes, the bound would still be violated by highly noncommutative black holes.  Rather than proposing novel bounds or searching over all systems for a minimum necessary $k=\underset{t\rightarrow \infty}{\lim} \frac{2M}{\pi \dot{S}}$ (giving the weakest bound on $\dot{S}$) to be the true proportionality in $\mathcal{C}=k S$, we suspect that the precise proportionality cannot be universally generalized between systems, at least under the established conventions for computing the WDW action.

\nocite{Brown:2016wib}
\nocite{Fischler:2000bp}
\nocite{Fischler:2013gsa}
\nocite{Hashimoto:2017fga}
\nocite{Karczmarek:2013xxa}

\bibliographystyle{JHEP} 
\bibliography{NCG} 

\end{document}